# Can 8-Year-Olds Understand Concepts of Einsteinian Physics?

Tejinder Kaur[1], Aishwarya Banavathu[1], David Blair[1], Rahul Choudhary[1], and Alex Foppolli[1]

[1] Department of Physics, The University of Western Australia, Perth, Western Australia

The decline of student interest in science in many developed countries has been demonstrated by the Program for International Student Assessment (PISA) scores since 2000. Furthermore, many studies in Australia have demonstrated a decline in the interest, motivation and performance of Australian students' across science, technology, engineering and mathematics (STEM) subjects over three decades. The Einstein-First project aims to reverse this trend by introducing children to Einsteinian concepts about space, time, atoms and gravity from an early age. This paper reports the first study on the ability of Year 3 students (ages 7 to 8) to conceptualise Einsteinian concepts. This intervention comprised of nine topics and a role-play, as well as a significant reinforcement learning of one class through in-class revision and an excursion to a science centre. Information was collected about students' conceptualisation and attitude towards learning modern science with pre- and post-intervention questionnaires. Video analysis of teacher-led Q&A sessions and feedback from the Class 1 teacher demonstrated a substantial uptake of Einsteinian concepts by students ages 7-8. A significant contrast in questionnaire results between the classes is interpreted in terms of teacher involvement and motivation. Results also indicate weaknesses in using questionnaire methods when observing younger students. This paper concludes that children aged 7-8 are not too young to learn key Einsteinian concepts, and the benefits of reinforcement indicate that longer programs are needed for ideas to be consolidated.

1. Introduction

The Program for International student Assessment (PISA) found that, since 2000 Australia, Finland, Iceland, Korea, the Netherlands, New Zealand, and Czechoslovakia, each had a decline in the mean performance of reading, mathematics and science (PISA, 2018). The performance of Australian students' undertaking compulsory education in science, technology, engineering and mathematics (STEM) subjects has declined over the last three decades (Rennie et al., 2001; Ainley et al., 2008; Lyons and Quinn, 2015). The factors underlying the reduced performance of students' in STEM include: i) a disinterest towards school science programs, that peaks in middle-school (<14 years of age) (Potvin & Hasni, 2014; Foppoli et al., 2019; Angell et al., 2004; Lovonen et al., 2017; Haeusler & Donovan, 2017); and, ii) a curriculum limited by an interpretation of Piaget's theory of cognitive development that implies that modern science concepts are out of the reach of children (Carey, 1985a; Vosaniadaou & Brewer 1987; Metz, 1995; 2008; 2009; National Research Council , 2007; Pitts et al., 2013; Haeusler & Donovan, 2017; Mandler, 1983; and, Gelman & Baillargeon 1983). The disengagement of students at the compulsory schooling levels, and the inability to entice school-leavers (Year 12: aged 16-17 years old) students to study STEM post-school (Tyler & Osborne, 2012; Securing Australia's Future report, pp 22-23), has resulted in the decline of university admissions and careers in STEM (Bakirei et al., 2017; Rennie et al., 2001; Lyons and Quinn 2010; Pitts et al., 2013).This led to Australia's Chief Scientist calling for further investment into STEM to secure continued economic, social, and cultural prosperity of Australia ( Australian Academy of Science 2015, 2016 ; Chief Scientist 2012).

A students' interest in science is the main determinant in pursuing a career in a scientific field. Maltese & Tai (2010) reported that student attitudes were developed before middle school. Long and Skamp (2008) found that positive attitude towards science is associated with students' enjoyment of practical science, new content, and unusual equipment. Contrary to Piaget's thinking that, children cannot handle big ideas of science until 14 years of age; James (1958) and Metz (1995,2008) showed that science curricula can be updated in accordance with childrens' development capabilities. Furthermore, research has shown that brain neuroplasticity is heightened in children between 5 to 10 years of age (Moreno et al, 2009).

Collectively, this suggests that the period of compulsory science – primary to year 10 may be the best time to introduce the foundational concepts on which modern science is based.

Internationally, physics educators have been trialling introduction of modern physics in school curricula (Kaur et al., 2017; Zahn & Kraus 2014; Henriksen et al., 2014; McKagan et al., 2008). Many authors have found that children as young as year 6 are capable of understanding modern science concepts of (Pitts et al., 2013; Kaur et al., 2017; Foppoli, 2019). In spite of these positive results, there is limited data with respect to primary school age students below Year 6 (Pitts, 2013).

This paper reports on a pilot study which is a part of a larger research project called Einstein-First which is developing an Einsteinian physics for students from Year 3 to Year 10 (Kaur et al., 2017). Einstein-First focusses on student learning concepts of the concepts and language of Einsteinian physics through role-playing and activity-based group learning using models and analogies.

The choice of year 3 for starting the Einsteinian curriculum was based on the success of earlier interventions and recommendations from teachers participating in our professional development programs. This paper presents the first test of the hypothesis that Einsteinian physics can be taught to younger children. First, we will provide a deeper review of the literature, and then present the methodology for our study. Finally, we will present the results of our study which contains two key components: a) the importance of a motivated teacher, and b) evidence for understanding of a variety of key concepts by a significant fraction of students. In the conclusion we give conditional 'Yes' to the title of the paper, but also include significant caveats.

   2.  **Literature Review**

Concepts of Einsteinian physics are not taught due to the assumption that they are too difficult for children to understand and also because the teachers have not learnt the content before. However, this attitude has been changing over time, partially because it is becoming more widely recognised that students are seeing more science and technology content on modern media which creates an increasing mismatch between students expectations of science and the science communicated at school (Haeusler & Donovan, 2017; Donovan & Venville, 2012; 2014).

The world has changed considerably since Piaget's time. Children are exposed to scientific and technological advances and the underlying concepts through their exposure to modern media (Haeusler & Donovan, 2017). Research in development psychology finds that cognitive development in children follows multiple diverse pathways, highlighting variability in cognitive development rather than the general view of *stages* by Piaget (Siegler 1996; Metz 1998; 2000; 2005; 2007). Furthermore, research suggests that mind functions by innate domain -specific modules, and that cognitive functions are highly specialised (Carey 1985a, 1985b; and Novak, 1977b). Thus, it has been argued that cognitive development in children should be approached within *domain*-specific theory, rather than developmental *stage*-dependent knowledge structure (Carey 1985a, 1985b; Novak, 1977b). Domain-specific theory shifts from a stage-dependent cohesive knowledge structure to more specialised domain specific knowledge structure.

The Einstein-First team uses role-play to teach concepts of Einsteinian Physics because it helps students look at occurrences of the universe from different perspectives (Dawson,1994). "*Role plays have been a recommended teaching strategy for their potential to make learning science more attractive to students who have become disenchanted with school science and female students in particular*" (Hildebrand, 1989). Ladrousse (1989) has underlined the many advantages for conducting role-plays: (a) it encourages students to create their own reality; (b) develops students' ability to interact with other people; (c) increases student motivation; (d) allows students to bring their experiences into the classroom; (e) helps identify misunderstanding and (f) provides shy students with a mask to participate, allowing greater student participation. From our point of view, role plays are a powerful way to cement modern terminology such as photon, atom and curved space.

Feedback is an integral part in developing content, it helps bridge gap between actual performance and desired outcomes (Carless, 2006). Given the importance of feedback in learning processes, it is also vital to facilitate discourse among participants in learning communities (Evans, 2013; Elkhouzai, 2016). This is especially well suited to group learning situations.

This paper includes discourse analysis of Year 3 students' interaction with the intervention presenter via video analysis. The value of video analysis has been to examine teaching and learning environment offers a means of close documentation and presents an avenue for interaction analysis and discourse analysis (Derry et al., 2010). It helps provide researchers powerful "microscopes" which provide substantial interactional details for comprehensive analysis (Derry et al., 2010). According to Richard and Lockhart (1999), interactive feedback is a specified approach to extend or transform a students' answer. In these terms, Discourse analysis is particularly a salient point for research in science education (Scott and Mortimer 2005).

In this paper, the video analysis provides evidence that complements the questionnaires which we found had limitations for such young students.

*2.1 Teacher Effectiveness*
Researchers have claimed that there is a relationship between teacher effectiveness and students' achievement. Effective teaching has been characterized in terms of specific practices such as employing systematic teaching procedures (Kemp & Hall, 1992), spending more time working with small groups of students (Taylor et al.,1999) and reinforcement learning (RL) (Chi et al., 2011).

Earlier exposure to modern STEM ideas through mass media makes children curious about modern technology and relevant discoveries, such as discovery of gravitational waves, black holes, communication systems, lasers, space exploration, nuclear power, renewable energy sources and the environment around them. Thus, mass-media acquired knowledge, even if rudimentary, is sufficient to ensure that students recognise that what they learning in school is, *"old stuff"*.

The traditional Newtonian paradigms taught in school, students' self-acquired interest in modern science concepts has not yet been matched with school curriculums. Students' interest, motivation, and attitude towards science declines sharply by 14 years of age because they are taught outdated concepts making it difficult for them to relate with reality (Potvin & Hasni, 2014). Several authors have emphasised that the "current" science curricula is laden with decontextualized content that is irrelevant to the experiences of the students that study them (Rennie et al., 2001; Pitts et al., 2013; Kaur et al., 2017). Einstein-First lessons ensure that lessons deal with topics that are relevant to the interest of students to help keep up their interest in science.

*2.2 Conflicts between the Newtonian and Einsteinian Paradigms*

The case for introducing Einsteinian physics depends crucially on recognition of the paradigm conflict between the core concepts of Newtonian physics are defined by concepts of Euclidian geometry, in which space is an absolute conceptual grid and time is absolute and unchangeable. In contrast, Einsteinian physics, which today is universally accepted as our best description of physical reality, space is deformed by matter and time changes with proximity to masses and relative speed and Gravity on Earth is caused predominantly by the stretching of time. Einsteinian physics also describes light as a stream of energy quanta which have a universal speed, and a frequency like a wave, but momentum and energy like that of a particle.

Because Einsteinian physics directly contradicts the core concepts of Newtonian physics, the early learning of Newtonian physics presents a significant obstacle to later learning of the modern paradigm. Moreover, modern physics concepts that are particularly attractive to young people, like black holes and lasers, cannot be discussed except in Einsteinian terms. The Einstein-First curriculum is designed to prevent cognitive confusion by introducing the Einsteinian concepts from an early age. The Newtonian approximation can easily be understood as a useful approximations (Shabajee and Postlethwaite, 2004; Pitts, 2013; Kaur, 2017).

Research about primary school students' understanding of Einsteinian Physics is limited; the evidence base comprises of survey data (Pitts et al., 2013; Kaur et al., 2017; Foppoli, 2019; Dua et al., 2020). In Australia the attitude and conceptual outcomes of Einsteinian physics have been described for Year 6 to Year 10 students as part of the Einstein-First project (Pitts et al., 2013; Kaur et al., 2017; Foppoli, 2019; Dua et al., 2020). Pitts' exploratory, single-school cohort study of Year 6 students in Australia reported that children did not consider themselves too young to learn Einsteinian physics concepts. Foppoli (2019) showed strong endorsement by students and teachers for introduction of Einsteinian concepts delivered via activity-based learning. Einstein-First team members report on the ease with which students accept the modern paradigm, but also report that the difficulty is to enable teachers to grasp the modern paradigm.

*2.3 Einstein–First*

The Einstein–First Project is an educational program founded by physicists and educators at the University of Western Australia. The project has developed an overarching pedagogy to resolve crucial teaching contradictions in primary and middle school science curriculum. The project has developed models and analogies to teach concepts and language of Einsteinian physics intuitively from an early age, thus giving students the best understanding of the physical universe. The project has developed courses, lesson plans and trains teachers to deliver Einsteinian physics lessons from Year 3 (7-8 years of age) to Year 10 (15-16 years of age). The goal of this project is to enable implementation of an Einsteinian physics-based curriculum worldwide to help alter students' ambivalent or negative attitudes towards STEM.

## 3. Methodology

In this study, pre- and post-questionnaires were used to assess students knowledge and attitude before and after the program. At the conclusion of the lesson, a video was recorded in which the presenter quizzed class 1 on their ability to verbalise Einsteinian notions.

*3.1 The Intervention*

The study was undertaken with two Year 3 classes at one of Perth's Catholic schools. Sixty students aged seven to eight years old participated in a nine-lesson program taught by three PhD students. The teacher of Class 1 actively participated in the program. However, because the class 2 teacher was on medical leave. The presenters were assisted in delivering the program by relief teachers.

*3.2 Nine In-Class Sessions*

In this study, nine lessons were delivered to both classes. Each 60-minute lesson included a 20-minute introduction to the idea via PowerPoint slides, a 20-minute activity in small groups, and a final 20-minute for worksheets and/or lesson-related discussions. Both classes were exposed to the same content.

In several instances the presenter could not complete the topic in the allocated time, often because the children were engrossed in activities, and asked for more time. In these instances, the lessons were completed the next day. The program covered the topics of space, time, light and gravity. The lesson plan of nine topics covered by the presenter is presented in Table 1.

**Table 1: Nine topics covered in the Einstein-First program**

| Topics | Description / Activities |
|---|---|
| **The History of Light** | The program was started with a role play called 'History of light'. Students played the roles of different scientists such as Einstein, Newton, Hertz, Feynman by wearing costumes such as Einstein's wig. A setup of old laboratories was created in the classroom. In the role play, they read about how the concepts of light developed over the time and the views of different scientists about it. Students learnt how scientists first thought of light as a wave and how the description changed to photons with time. Students really enjoyed playing the roles of different scientists and did the play multiple times because every student wanted to have a turn to play the role of any scientist. The script of the role play is in Appendix 1. |
| **Curved space** | After the role-play, the next lesson was based on the space-time simulator where students performed several experiments to understand how space is curved with mass. The lesson was started with a question – 'what is space?'. Students' views about space were then collected. One of the misconceptions among the students was that space is something which is far away from us. They believed if we talk about space, it means we are talking about outer space, about the universe. It was made clear to them that space is everywhere, there is space between our hands, space in the room, space between our two ears.<br><br>To consolidate their views, students rolled different sized marbles on the spacetime simulator setup. They started the activity by adding one marble on the setup and observing the curved surface. After which they gently added another ball and observed that suddenly gravity appeared and both balls get attracted toward each other. By adding more and more marbles, they created a big curvature in the spacetime simulator. Students also understood how the Sun was created and gravity is the curvature of spacetime. Students summarise the activities through John Wheeler's phrase:<br>Matter tells spacetime how to curve;<br>Spacetime tells matter how to move |
| **Mapping Curved Space** | To teach the concept of light deflection, students used pull-back cars on the spacetime simulator. These pull-back cars represent photons from a distant star. Students released the cars and observed that, in the beginning, two cars were parallel, as soon as they approached the central mass, their paths crossed, |

| | which showed that light is deflected near strong gravitational fields. We also used this analogy to teach the concept of parallel lines where we showed that parallel lines can meet if the surface is curved. |
|---|---|
| **Curved space geometry** | Students performed experimental geometry on works to determine how geometry on a flat piece of paper differs from geometry on a curved surface. Students created various sized triangles on the work surface and observed how the sum of the angles of each triangle varies with its size. |
| **Photons and Nerf gun photography** | Students learnt the concept of photons with the help of Nerf guns where Nerf guns represent light sources, and the bullets represent photons. Students photographed each other against a whiteboard with Nerf gun bullets, where bullets stuck to the whiteboard and created silhouette images. By doing this activity, students learnt photons have momentum and that is how real cameras register photons on the CCD device. |
| **Photoelectric effect** | To understand the photoelectric effect, students knocked out ping-pong balls, representing electrons, from saucer of different depths with Nerf gun bullets. They found out how electrons got knocked out with analogue photons. |
| **Soap Film Interference** | To understand the wave properties of light, students did a few experiments with low power lasers. Students shone laser light on a soap film created on a circular wire, which produced the inference pattern of dark and bright fringes. |
| **Measuring Hair Widths** | In this lesson, students observed the interference pattern by shining a laser light on the hair. They observed how light got diffracted around the hair and produced dark and bright fringes. |
| **Math of Arrows** | Students were handed two arrows and were asked to align them in a such a way that 1+1 makes 2 and 1+1 = 0. Students immediately picked up the idea and put two arrows from head to tail in one direction to make 1+1 = 2 and put two arrows in a opposite direction to make 1-1=0. By doing this activity, students grasped the idea of where the probability of arrival of photons is high to make bright fringes and where the probability of the arrival of photons is low to make dark fringes. Students summarise the light activities with the following phrase: "Everything has bulletiness, everything has waviness; Bullets follow the maths of numbers; waves follow the maths of arrow". |

### 3.3 Data collection

To assess students' conceptual understanding towards Einsteinian physics and their attitude towards science, four questionnaires were developed. These questionnaires were carefully designed under the guidance of physicists and educators. These questionnaires have been used in previous trials and the method used to create these questionnaires is described in another research paper written by the authors of this paper (Kaur et al. 2018).

#### 3.3.1 Conceptual pre/post- questionnaires

To assess students' conceptual understanding before and after the programme, conceptual pre- and post-tests were designed. Both tests were identical. There were eight questions, including those requiring short answers and multiple-choice answers. Students were given 15 minutes to complete these tests. The questionnaire is given below in Table 2.

#### 3.3.2 Attitude pre/post questionnaires

To assess students' attitude towards science before and after the programme, pre- and post-attitudinal questionnaires were prepared. Both questionnaires had 10 identical questions based on five Likert scale responses - strongly disagree to strongly agree. Students were given 10 minutes to complete this questionnaire. The questionnaire is given below in Table 3.

### 3.4 Data analysis

The student responses to knowledge questions were marked as,

 0 = No response, incorrect or did not use Einsteinian language;
 1 = Partially correct and consistent with Einsteinian; or

    2    = Correct and consistent with Einsteinian language.

Students attitudinal responses were based in Likert scale items and were recorded as

  1= strongly disagree, 2= disagree, 0 = neutral, 4 = agree and 5 = strongly agree

*3.5 Excursion Event (Class 1 Only)*

Following the Nine *In-Class* topics, the students of Class 1 participated in a full-day excursion to a public science centre called the Gravity Discovery Centre (GDC). During this excursion they participated in a role-play activity. Students separately rehearsed the role-play with assistance of school drama teacher before performing it at GDC.

*"How to make Gold"* role play is based on the 2017 observation of gravitational waves from a pair of coalescing neutron stars. The event encompasses the entire breadth of Einsteinian physics from the structure of atoms to curved space and black holes. Its discovery followed an astonishing sequence of scientific predictions made throughout the 20$^{th}$ century, and solved the mystery of how gold is made in the universe. The play explores the process by which prominent physicists including Isaac Newton, Albert Einstein, Robert Oppenheimer, Jocelyn Bell and Tsvi Piran discovered how to make gold. Five children were selected play the role of the physicists and another five children played the role of students. The role-play culminates in the idea that when "neutron stars collide neutron matter is released and explodes, in a process that creates gold and flings it out into Space". The full role-play can be accessed on https://www.einsteinianphysics.com/role-plays/.

*3.6 Video Analysis of presenter-led Question Session*

To complement questionnaires, the presenter conducted a video-recorded oral question-and-answer session which was conducted on the final day of intervention with Class 1. The purpose was to get a sense of the general level of knowledge and conceptual understanding and student ability to communicate Einsteinian concepts. The oral responses of the students were analysed, and assessed for the enthusiasm and clarity of their answers. The presenter recorded the classroom interaction using a mobile video camera for the duration of the question-and-answer session only. The participants were 'blinded' to the reasons why they were being observed in an effort to minimise 'participation bias'. This video analysis used an analytical framework classroom discourse to describe the interactions, which factors in the content, type of utterance, thinking and interaction pattern of the students. The authors manually coded and transcribed the video and assigned i) movement a form of utterance, ii) initiation (I), and; iii) response (R) (see Table 4). The presenter started the questionnaire session with an open-ended question, *"What do you remember from the classes?"*

## 4. Results and Discussion

This section presents research findings and discusses the outcomes of the intervention. We begin with the findings of students' knowledge assessments, followed by their attitudes towards science. Then we discuss the video analysis and analysis of teacher feedback.

*4.1 Pre- and Post-Intervention Knowledge Test Results*

The outcomes of this study, which involved two Year 3 classes of 54 students in one of the primary schools in Western Australia, indicated that there were demonstrable and gains in students' understanding of Einsteinian physics concepts.

The following histograms illustrate the outcomes of two different classes of students' knowledge. Pre-test scores are zero because students responded in Newtonian language to the pre-test questions. Section 4.2 contains typical examples. Students post-test results are listed in ascending order.

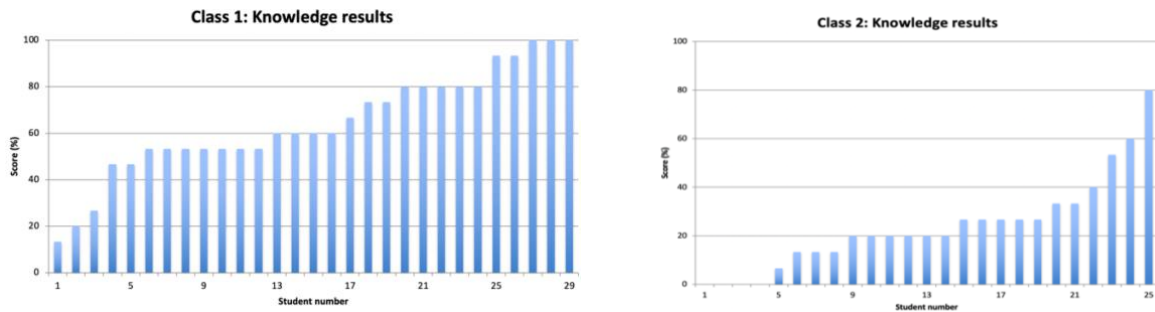

Figure 1: (a) Results of the Class 1 post-intervention knowledge test, arranged in increasing order. (b) Results of the Class 2 post-intervention knowledge test, arranged in increasing order

The post-test results for twenty-nine students in class 1 are shown in Figure 1a, whereas the post-test results for twenty-five students in class 2 are shown in Figure 1b. Class 1 students outperformed class 2 on the post-test, as can be seen from the graphs. Class 1 had 17 students who scored at least 60%, whereas class 2 had only two students who scored 60%. The majority of students in class 2 scored below 40%, while just 10% of students in class 1 scored below 40%. There are three class 1 students who achieved 100% score on their post-test results. Class 1 has an average of 64%, whereas Class 2 has an average of just 24%. Their average scores are very different. Later in the paper, the various explanations for the disparity in the results of two classes are discussed.

### 4.2 Analysis According to Each Question

A typical answer given by students to each question are presented in this section. The results are shown in Table 2.

### 4.2.1 A question on Space: KQ1

KQ1 was created to assess students knowledge of space. The concept was explained using the spacetime simulator activity outlined in Table 1. Students were taught that space is similar to a stretchable, bending fabric; we don't know what space is, but we do know how to measure it. John Wheeler's phrase *"Matter tells spacetime how to curve, spacetime tells matter how to move"* was used to reinforce the concept.

KQ1 – "Space? What is it? What is it like? Please give your ideas".
In the pre-test, students described space as something far away from us, e.g., "Outer Space".
Some of the pre- intervention student responses are listed below: -
- "It is dark and full of stars", "Space is above Earth and quiet dark", "It includes Sun, Moon, Earth and Mars", "Space is where there is not much gravity and …has nine planets", and "Space is an area that does not have anything in it".

KQ1 post intervention results – 21% of Class 1 and 12 % of Class 2 answered correctly and 31% of Class 1 and 40% of Class 2 answered it partially correct.

### 4.2.2 Questions on Light: KQ2, KQ4, KQ8

Questions KQ 2, 4, and 8 were designed for testing students learning about light. Light comes as photons was explained using the Nerf gun activity described in Table 1. The phrase "Everything has bulletiness and everything has waviness" was used to emphasise the dual nature of light. The following three questions were asked to test students understanding of photons, which have the ability to push things and nothing can go faster than the speed of light.

KQ2- "Light: What is it? What is it like? Please give your ideas".

KQ4- "Can Light push things? Circle Yes or No. Please give your answer".

KQ8 – "Some airplanes can go faster than Sound? If you had a super-powerful rocket, could you go faster than light? Please explain".

Students in their pre- test described light as something that helps us see objects or that it has warmth. Some of the pre- intervention student responses are listed below: -
- "It is something you use to see", "Light is something bright, it can shine and can warm you up", "Light is when everything you can see and it is hot", "It is bright shine like the sun".
- "No, because it is very light and it is not solid", and "Light cannot push things because it doesn't".
- "Yes, because rockets are powerful", and "Yes, it can go faster, because the rocket is powerful".

None of the students mentioned photons in their responses. Students mentioned that rockets are powerful and could travel faster than light speed. Students were unaware that light has momentum and thus the power to push objects. However, after participating in Nerf gun activities, kids were able to conceptualise light in terms of photons, which have the potential to push objects. In the post-test, 41% of students in class 1 and 16% of students in class 2 correctly answered KQ2, whereas 17% of students in class 1 and 16% of students in class 2 answered it partly correctly. In response to KQ4, 21% of Class 1 and 12% of Class 2 correctly answered, while 45% of Class 1 and 60% of Class 2 partially answered. KQ8, nothing can go faster than the speed of light, 48% of Class 1 and 24% of Class 2 correctly answered, whereas 7% of Class 1 and 4% of Class 2 were only able to just say 'No' in their responses but were unable to explain their responses.

*4.2.3 A Question on Gravity: KQ3*

KQ3 was designed for testing students learning about Gravity. Students did the activities on the spacetime simulator to understand the concept of Einsteinian gravity which is the curvature of spacetime.

KQ3- "What is Gravity?".

In the pre-test, students described gravity as an instantaneous force which holds us on the ground. Some of the pre-intervention student responses are listed below: -
- "Gravity is something that pulls you down and make sure you don't float and always come down", "...Whatever goes up must come down", "It is a place that does not have air", and "Gravity is the things that pulls us onto the ground".

After attending the program, several students were able to answer in Einsteinian language. It was correctly answered by 24% of Class 1 and 4% of Class 2. When the results were discussed with the class teacher, he mentioned that students had difficulty to understand the concept of gravity.

*4.2.4 Questions on Curved Space Geometry: KQ5, KQ6*

KQ5 and KQ6 were designed for testing students learning about Curved Space Geometry. Students drew triangles on woks and tested the parallel lines hypothesis on the Spacetime simulator. Students had a great difficulty measuring angles because they hadn't been taught how.

KQ5- "How could you tell if the ruler is straight? Give two or more ways that you could".

KQ6- "The straight lines in the diagram are called parallel lines. If you extend the lines so that each one stayed perfectly straight, could they ever meet?".

Some of the pre-intervention student responses are listed below: -
- "Rulers are straight, and they look straight", "It looks straight as it does not curve at all'.
- "No because they are parallel", "Parallel cannot go curved so it is straight", "Parallel lines never meet they just keep going straight" and, "No, because they will stay straight and keep extending".

In the lessons, students learnt that we could tell of the ruler is straight by comparing it with light or by sighting and comparing with other straight things and for KQ6, they learnt that parallel lines can cross each other because of the Curved Spacetime.

After the program, 38% of Class 1 and 32% of Class 2 answered KQ5 correctly and 24% of Class 1 and 24 % of Class 2 answered it partially correct. For KQ6, 10% of Class 1 and 8% of Class 2 answered correctly and 10% of Class 1 and 28 % of Class 2 answered it partially correct.

The pullback cars activity explained two concepts; i) how light gets deflected near strong gravitation fields and; ii) how parallel cars represent two parallel lines crossing their path when approaching the gravitational field. The results could justify that Year 3 students could not relate KQ5 and KQ6 to the activities they performed.

*4.2.5 A Question on Time: KQ7*

KQ7 was designed for testing students learning about the concept of time. The concept was taught using PowerPoint slides and YouTube videos.

KQ7- "Is time different at the top of a building compared with the ground floor?".

Some of the pre-intervention student responses are listed below: -
- "No, because time is minutes and if you were anywhere else it would be the same"," Time is different when you are somewhere else in the world", and "Time is always the same even though it feels longer when you are bored".

After the program, students mentioned that, yes, times runs slower near to stronger gravitational force. Time is slower at the bottom and faster at the top. After the program, 14% of Class 1 and 32% of Class 2 answered correctly and 34% of Class 1 and 56 % of Class 2 answered 'Yes' but couldn't justify their responses.

**Table 2: Results of Knowledge Questions Post- intervention for Class 1 and Class 2**

| Knowledge Question (KQ) | Questions asked in the pre-/post test | Class 1 | | Class 2 | |
|---|---|---|---|---|---|
| | | Correctly (%) | Partially (%) | Correctly (%) | Partially (%) |
| 1 | Space: What is it? What is it like? Please give your ideas | 21 | 32 | 12 | 40 |
| 2 | Light: What is it? What is it like? Please give your ideas? | 41 | 17 | 16 | 16 |
| 4 | Can light push things? Circle Yes or No. Please give a reason for your answer. | 21 | 45 | 12 | 60 |
| 8 | Some airplanes can go faster than sound? If you had a super-powerful rocket, could you go faster than light? Please explain. | 48 | 7 | 24 | 4 |
| 3 | What is Gravity? | 24 | 3 | 4 | 0 |
| 5 | How could you tell if a ruler is straight? Give two or more ways that you could tell. | 38 | 24 | 32 | 24 |
| 6 | The straight lines in the diagram are called parallel lines. If you extend the lines so that each one stayed perfectly straight, could they ever meet? | 10 | 10 | 8 | 28 |
| 7 | Is time different at the top of a tall building compared with ground floor? Circle Yes or No. Please give a reason for your answer. | 14 | 34 | 32 | 56 |

*4.3 How does the Einstein-First program effect students' interest in engaging with science?*

Student attitude towards science was administered by an attitudinal questionnaire which is given below in Table 3. As mentioned in the earlier section, the Attitudinal Questionnaire (AQ) employed a five - point Likert scale and the students were asked to rate identical questions before and after the intervention.

**Table 3: Attitudinal Questionnaires (Pre- and Post-Intervention)**

| Category 1 | | Attitude towards learning modern concepts |
|---|---|---|
| | AQ1 | I think gravity and light are interesting topics. |
| | AQ2 | I think space and time are interesting topics. |
| | AQ3 | I do not enjoy learning new concepts and ideas. |
| **Category 2** | | **Attitude towards learning through hands-on activities** |
| | AQ4 | I prefer to learn science by doing activities with other students. |
| | AQ5 | I do not enjoy doing science activities. |
| | AQ7 | I would rather learn from other people than do activities. |
| **Category 3** | | **Attitude towards science** |
| | AQ6 | I would like to be a scientist when I grow up. |
| | AQ8 | Science is only for smart people. |
| | AQ9 | I do not have much interest in science. |
| | AQ10 | I like science more than any other school work. |
| | AQ11 | Science does not help me to understand the world. |

The responses of these questions within each category were combined to assess how many students were interested to learn the modern concepts.

*Category 1: Attitude towards learning modern concepts*

The Einstein-First program introduced students to intriguing topics that they previously had not considered to be classroom science. The presenter was befuddled on the first day of the program by numerous questions such as "Are we going to learn about black holes?" and "Will you teach us about the universe?" When the presenter asked, "Could you please name a famous scientist?" "Albert Einstein" was the majority of responses. "What made Einstein famous?" "$E= mc^2$" was the response. "What does this imply?" "don't know" was the response. We discovered that while these students were familiar with modern science which they learned through documentaries, news, and YouTube videos, none of these topics were taught at their year levels or included in the science curriculum.

Questions AQ1, AQ2, and AQ3 were created to ascertain student attitudes towards modern topics. The results suggested that 65% of class 1 students and 80% of class 2 students disagreed with the statement "I dislike learning new concepts and ideas," and that most students considered space time, gravity, and light to be interesting.

*Category 2: Attitude towards learning through hands-on activities*

The students gained an understanding of modern concepts such as space, time, gravity, and light using hands-on activities listed in Table 1. Questions AQ4, AQ5, and AQ6 were meant to collect information on the students' attitudes towards active learning. During the program, students appeared to be totally absorbed in the activities, evident by their excitement and desire to participate. Additionally, the students participated in a role play to learn about the history of light. The presenter had a difficult time choosing seven children from a class of thirty to play Einstein, Newton, and other scientists. As a result, the play was repeated four times to ensure that all students were accommodated. Each day, when the session concluded, kids were eager to learn what their next lesson will be and what activity they would participate in. The responses to AQ 5 and AQ 7 demonstrated the students' interest in learning through activities. 80% of students disagreed with the statement "I do not enjoy doing science activities," and 70% preferred to learn through activities rather than from another person.

Class 1 received additional time to complete the activities, and the teacher reiterated each idea. Responses to AQ4 "I prefer to learn by doing" indicated that class 1 students felt confident in their ability to do the activities independently, with just 34% agreeing with the statement. However, class 2 did not receive extra time for activities and had no reinforcement of the concepts; 64% of students from Class 2 agreed with this statement.

*Category 3: Attitude towards science*

This category summarised how students' attitudes towards science learning changed as a result of their participation in the programme. Students expressed their opinions by responding to five questions AQ6, AQ8, AQ9, AQ10, and AQ11. Most students demonstrated a favourable attitude. In response to AQs 8 and 9, 88 percent of students disagreed that science is only for smart people and 80 percent agreed that science helps them to understand the world. The majority of students, 80 percent, stated that they were interested in science and preferred it to other schoolwork. Their curiosity in science was stimulated by the Einstein-First hands-on activities. AQ6, asked the about scientific careers, by uncovering student perceptions of the relevance of science to their future careers. Careers in science were not specifically covered by the program, so it was difficult for students of this age group to talk about future careers. Only 20% of students agreed that they wished to be a scientist when they grew up. The Einstein-First program could help to spark students' interest in science, and they might pick science as a subject in subsequent school years.

*4.4 Video analysis of teacher led questionnaire*

Video analysis of a teacher-led question-and-answer session demonstrated a high proportion of students captured during the recorded session reacted positively to the presenter's questions. Students' would generally move (hands up) and respond (provide an answer when asked by a teacher), rather than initiate (blurt out an answer without being called to do so).

Despite partial obstruction of vision, the data demonstrates that children were responsive and enthusiastic in answering the questions. The questions used by the presenter were supplemented with interactive feedback to help extend or transform students' answers. At face value the teacher led questionnaire might only give opportunity to students that raise their hands first, however the body language, the quickness to raise hands, confidence and enthusiasm coded in a discourse analysis are difficult to quantify but they are consistent with KQ.

**Table 4: Video Analysis of Teacher-led Questionnaire**

| Presenter-Led Questions | Move n (%) | Initiation n (%) | Respond n (%) | Any reaction n (%) |
|---|---|---|---|---|
| So, what do you all remember from all the class? (n=14*) | 10 (71.4) | 3 (21.4) | 1 (7.1) | 14 (100) |
| Time can be different in different places. Great! what is an example on how time can change? Who remembers? (n=13*) | 7 (53.8) | 5 (30.8) | 2 (15.4) | 12 (85.7) |
| What else did we learn about? (n=14*) | 6 (42.9) | 1 (7.1) | 5 (35.7) | 12 (85.7) |
| Nice, really good! Okay let's get someone else to say it.. may be a girl. (n=14*) | 5 (35.7) | 5 (35.7) | 3 (21.4) | 13 (92.9) |
| Light travels in photons. Great! Who can best explain what is a photon? (n=17) | 7 (43.8) | 3 (17.6) | 3 (17.6) | 13 (76.5) |
| Almost. It's got energy. (n=14*) | 4 (28.6) | 0 (0.0) | 9 (64.3) | 13 (92.9) |
| Not matter. But it begins with an 'M'. (n=29*) | 7 (24.1) | 4 (13.8) | 15 (52.7) | 26 (89.7) |
| Right! Now how do we get 1+1=2? (n=29*) | 5 (17.2) | 1 (3.4) | 18 (62.1) | 24 (82.8) |
| * n=assessor had unobstructed vision of children in the video to analyse student responses. | | | | |

## 5. Teacher feedback analysis

The regular teacher of Class 1 provided favourable feedback on the appropriateness of the delivery of the Einstein-First project to Year 3 students. Favourable feedback cited that students were excited to learn about physics, the concepts were taught at an appropriate level, and that the intervention was a great benefit to the students. The teacher provided constructive criticism, which focussed on the presentation skills (presenter spoke too long), content (some of the content was too difficult), and testing (the questions were too hard).

*Overall Intervention Feedback*

"We were very lucky to have the Science Team from the Physics Department at UWA come and teach us some Einsteinian Physics. The Year 3 classes were excited to learn about physics and the concepts covered. The program had some difficult concepts that were broken down to the level where the children could understand the principles being taught."

*Content Feedback*

"The experiments with the space/time simulation were very engaging and the children loved rolling marbles and ball bearings onto the fabric simulating planets and gravity. Using a Nerf gun to simulate light particles was a great way for younger children to understand how light moves in particles and why/how a shadow is formed."

*Presenter Feedback*

The presenters used hands-on activities and the smartboard to teach Einsteinian concepts. Sometimes the presenters spoke for too long at one time and some of the children were losing focus. Some of the content was above the understanding of the class but most of them seemed to understand most of what was covered."

*Questionnaire Appropriateness Feedback*

"In the final examining of the children's knowledge the written questions were too hard for some of them to understand and answer, however they could verbally answer questions accurately."

*Teachers' Attitude Feedback*

"All-in-all I loved it and think the kids will greatly benefit from having had UWA physicists in to teach them Einsteinian Physics."

## 6. Conclusion & Discussion

The purpose of this study was to determine whether lower primary school students are capable of understanding Einsteinian physics in the classroom We assessed students' conceptual understanding of Einsteinian physics and the impact of the program on their attitudes towards science through knowledge and attitude questionnaires. We will discuss the significance of our findings in this part as well as the methodological issues and limitations of our research design.

This was the only study in Einsteinian physics education that was undertaken with lower primary school students. Therefore, our findings contribute to the Einsteinian physics education by providing detailed descriptions of how lower primary school students perceive Einsteinian physics concepts.

In general, we discovered that using a hands-on approach to present Einsteinian physics concepts benefited students in comprehending the physical reality of the universe and sparking their interest in science. Einsteinian physics covers the notions of curved spacetime, experimental geometry on curved surfaces and photons, which are all fundamental concepts for students to understand their favourite topics such as black holes. Students can relate to Einsteinian physics since it is directly applicable to our daily lives. Another study indicated that students' interest in science can be improved by incorporating Einsteinian physics topics of black holes, gravitational waves, or dark matter into their classrooms (Woithe & Kersting, 2021)

We found that while lower primary school students could cope with Einsteinian physics concepts, each notion required additional time to comprehend properly. We compressed the whole program into ten days, which is one of the program's limitations. At this age of students, reinforcement is critical. In this study, class 1 teacher reinforced the learning through repetition, but this was not done for Class 2.

We also found that class 1 did better than the class 2 and there are several assumptions on why one did better than the other. There were different teachers in both classes, although the presenter was the same and the same Einsteinian content was presented to both classes.

Class 1 teacher has a science background whereas class 2 teacher was a relief teacher because their regular teacher was on medical leave. Class 1 teacher revised the content the day after the presentation to strengthen student understanding of the concepts and this did not happen with class 2. As for the team, this was their first time delivering the program to young students of 7 – 8 years old so a few of their assumptions regarding students' existing knowledge were wrong. For example, the team had one lesson on curved space geometry where students drew triangles and measured the angles with protractors. The team assumed that the students knew how to measure the angles with protractors, but the students had not learnt that yet. Class 1 teacher had some extra lessons to teach on how to measure the angles which did not occur with the other class. Class 2 never repeated any activities or lessons after the presenter left.

Class 1 teacher also realised that the knowledge pre and post tests were difficult for the students to answer. The teacher said it would be good if the team used multiple-choice questions instead. Class 1 teacher said the students knew the answers, but it was difficult for them to articulate their answers and put them into writing.

Also, class 1 teacher used the strategy to write a short paragraph on what they have learnt in the lesson and picked up immediately where the students were lacking, or which concept did not come across accurately. The teacher explained that again to make sure they understood the concept. Class 1 teacher was also aware what the students were going to do beforehand. The teacher prepared the students on what they were going to learn to make sure they could follow what the presenter was going to teach. This was the extra advantage for class 1 and class 2 did not get these opportunities.

On the last day of the program, the team recorded a video where they asked a few questions to the students. The team found that the students gave very good responses, and many hands were up when the team was asking questions. The students easily summarised the concepts by using two phrases about spacetime and bulletiness. The students' responses included the phrase momentum, photons, which was previously thought to be beyond the students' comprehension at this age. The transcript of the video is given in appendix 2.

The team had an enrichment lesson on neutron stars and how to make gold. The students did a role play in the class with the help of a drama teacher. An interesting outcome of the program was that observers were so impressed by the role play performance that the Class 1 students were invited to present the play at an official opening event at a public science education centre before a distinguished audience. This took place after the course evaluation, so the performance did not contribute to our assessment in this paper. The video of this role play is available on the website www.einsteinianphysics.com .

Additionally, our data found that both genders had an equal interest in learning science in lower primary school. The program includes stories about male and female scientists who exemplified extraordinary acts of creativity and inventiveness, demonstrating that science is not only for boys, but for everyone. Previous research indicates that girls were more likely than boys to reject science as a career choice due to their inability to envision themselves as scientists (Lyons & Quinn, 2010). By including both male and female scientists in the role-play, both genders can see themselves as scientists. In this study, we discovered that girls are equally enthusiastic and interested in participating in activities and acquiring modern concepts. In summary, primary school students demonstrate a strong interest in science and no gender influence was observed.

## 7. Limitations

This section discusses the limitations of our study considering its research strategy and methodology. As mentioned in previous sections, this was the first study to examine primary school students' responses to Einsteinian physics concepts. The team had validated its teaching approaches through multiple studies with upper high school students. To implement this program, the team relied on a suggestion made by one of the teachers attending the professional development course that Year 3 students could grasp Einsteinian physics concepts. Soon after, another teacher asked the team to conduct a trial run of the program with his Year 3 students. The team jumped at the opportunity and chose to complete the program during the same term that was coming to an end.

One of the study's weaknesses was that the researchers chose to trial the same program they regularly use with high school students. Concepts have been broken down to make them more accessible to Year 3 students. The presenter lacked experience teaching Einsteinian physics to 7- or 8-year-old students and made no attempt to contact the teacher when designing the program and evaluation assessments. The program was not simplified sufficiently as per their age level. The team did not enquire about students' prior knowledge when designing the lesson plans and encountered some issues delivering a few ideas. For instance, a one-hour lesson was created to teach students experimental geometry on curved surfaces. While delivering that lesson, the presenter discovered that students had not yet learned angle measurement. As a result, the planned one-hour lesson took a week to fully comprehend. The teacher in class 1 made extraordinary efforts to ensure that students grasped the concept of geometry on curved surfaces and assisted class 2 as well. Both class teachers had never attended an Einstein-First teacher session and were unfamiliar with the content. After delivering the first two sessions, the class 1 teacher saw that students were failing to comprehend Einsteinian physics vocabulary and began inquiring about the upcoming lesson to ensure students were adequately prepared to comprehend all of the presenter's terms.

Another limitation of the study was of using only quantitative approach to assess students understanding in Einsteinian physics concepts. Open ended questions were asked in the pre- and post-tests. At this age, students writing skills were not fully developed and found it difficult express their understanding in written form. The team did not use the multiple-choice questionnaire with the fear that students might guessed correct answers and it will be difficult to understand whether they really understood the concept of they just chose answers randomly. To validate whether it is appropriate to use multiple choice questionnaire, class 1 teacher trailed it. The teacher collected students' responses to a multiple-choice questionnaire and repeated the process after one week. As predicted, students chose different answers week later which validated that even multiple-choice answers were not good enough. At the end of the program, we asked a few questions to class 1 and found that students were perfectly answer every single question. Students were clearly able to explain momentum. This interview was not planned but proven to be an effective method to check students understanding. It was conducted with the whole class which limited the opportunities for every student to answer the questions.

Taken together, if Einsteinian physics is taught to students in a manner commensurate with their age and current level of education, then further iterations of the Einstein-First program could achieve better results in younger student groups. We hope that future research might build on our findings to take this understanding further.

**Acknowledgements**
This research was supported by a grant from the Australian Research Council, the Gravity Discovery Centre. We would like to acknowledge the contribution of all the project team members and, in particular, the teachers and students who participated in this program on voluntary basis.